\documentclass[english,11pt,aps,prd,a4paper,preprintnumbers,nofootinbib,superscriptaddress,notitlepage]{revtex4}
\usepackage[T1]{fontenc}
\usepackage[latin9]{inputenc}
\usepackage[usenames,dvipsnames]{color}  
\definecolor{darkred}{rgb}{0.6,0,0}
\definecolor{brown}{rgb}{0.59, 0.29, 0.0}

\definecolor{darkcyan}{RGB}{0, 111, 111}
\definecolor{dgreen}{rgb}{0,0.5,0}

\usepackage{hyperref}
\usepackage{ulem}
 
\newcommand {\ignore}[1]{}
\usepackage{amsmath,amssymb,mathtools,graphicx,fancyhdr}
\usepackage{multirow}
\usepackage{graphics, setspace}
\usepackage{comment}
\newcommand{\AddrAHEP}{AHEP Group, Institut de F\'{i}sica Corpuscular --
	C.S.I.C./Universitat de Val\`{e}ncia, Parc Cientific de Paterna.\\
	C/Catedr\'atico Jos\'e Beltr\'an, 2 E-46980 Paterna (Val\`{e}ncia) - SPAIN}

\newcommand{\AddrCinvestav}{Departamento de F\'{\i}sica, Centro de
	Investigaci{\'o}n y de Estudios Avanzados del IPN\\ Apdo. Postal
	14-740 07000 Ciudad de M\'exico, M\'exico}

\newcommand{\AddrTEC}{Tecnol\'ogico Nacional de M\'exico/ITS de Jerez,
	C.P. 99863 Zacatecas, M\'exico.}

\begin{document}

\title{Testing the non-unitarity of the leptonic mixing matrix  at FASER$\nu$ and FASER$\nu$2}

\author{Jes\'us Miguel Celestino-Ram\'irez}\email{jesus.celestino@cinvestav.mx}\affiliation{\AddrCinvestav}

\author{F. J. Escrihuela}\email{franesfe@alumni.uv.es}\affiliation{\AddrAHEP}

\author{L. J. Flores}\email{ljflores@jerez.tecnm.mx}\affiliation{\AddrTEC}

\author{O. G. Miranda}\email{omar.miranda@cinvestav.mx}\affiliation{\AddrCinvestav}

\begin{abstract}
  The FASER$\nu$ experiment has detected the first neutrino
  events coming from LHC. Near future high-statistic neutrino samples
  will allow us to search for new physics within the neutrino
  sector. Motivated by the forthcoming promising FASER$\nu$
  neutrino data, and its succesor, FASER$\nu$2, we study its
  potential for testing the unitarity of the neutrino lepton mixing
  matrix. Although it would be challenging for FASER$\nu$
  and FASER$\nu$2 to have strong constraints on
  this kind of new physics, we discuss its role in contributing to a
  future improved global analysis.
\end{abstract}


\maketitle

\section{Introduction}
The neutrino oscillations discovery tells us that neutrinos have a
small mass. Compared with other fundamental particles, the nonzero
neutrino mass and its smallness strongly suggest that the Standard
Model (SM) needs an extension to describe the neutrino oscillation
picture. Also, the SM needs a new mechanism or an explanation for the
mass degeneracy of the active neutrinos. An attempt to describe the
mass generation of neutrinos is the seesaw
mechanism~\cite{Minkowski:1977sc,Gell-Mann:1979vob,Yanagida:1979as,Mohapatra:1979ia,Schechter:1980gr}. The
type-I seesaw mechanism uses neutral heavy leptons (NHL), with
Majorana mass, as a messenger to transport mass to the light
neutrinos. Due to their heavy mass, the NHLs do not oscillate to
active neutrinos. However, their effects are contained in a submatrix
of the full $N\times N$ lepton-mixing matrix, with $N$ the number of
light plus heavy neutrino species. As a consequence, the 3$\times$3
mixing matrix of the light neutrino is non-unitary. In recent years,
many experiments have been used to test non-unitarity
effects~\cite{Gronau:1984ct,
  Nardi:1994iv,Atre:2009rg,Escrihuela:2016ube,Fernandez-Martinez:2016lgt,Blennow:2023mqx,Forero:2021azc,Denton_2022,Dutta:2019hmb}.
Among the new experiments expected to give further information about
neutrino interactions, we can consider the case of FASER, and more
especially FASER$\nu$, which will measure the neutrino cross-section
in a new energy window, making this experiment an exciting place to
study either Standard Model physics or beyond.  

In this work, we will
explore the non-unitarity sensitivity in the FASER$\nu$ and FASER$\nu$2
experiments. With different neutrino channels measured at high
energies, FASER$\nu$ will test non-unitarity effects in an
experimental setup different from any other experiment. Therefore,
this makes the study of a future non-unitary test at FASER$\nu$
interesting.  FASER$\nu$ experiment works at
100-1000~GeV~\cite{FASER:2019dxq}, and the momentum transfer for this
fixed target detector will be around $Q^2\sim (10~
\mbox{GeV})^2$~\cite{FASER:2019dxq}.  At this energy, it might be
possible to generate NHL for specific theories, like a linear or
inverse seesaw below the electroweak energy, for example, in the mass
range of GeV, as was studied in~\cite{Kling:2018wct}.
In this work, we will focus on a model-independent formalism for
non-unitarity~\cite{Escrihuela:2015wra}, valid for neutrino mass
eigenstates at high mass scales, above hundreds of GeV, and show the
sensitivity to the corresponding parameters.
A different study for the non-unitary case was done
previously~\cite{Aloni:2022ebm}. However, their approach is different
in terms of the theoretical description of non-unitarity as well as
in the study of other neutrino observables in their analysis.

The paper structure is: in section II we briefly review the non-unitarity formalism and the zero-distance approximation used in this work. In section III we show the statistical procedure that we follow to obtain the sensitivity in the non-unitarity formalism. The $\chi^2$ analysis and the sensitivity of the non-unitarity parameters are discussed in section IV. Finally, in section V, we talk about the conclusions and perspectives. 

\section{Non-unitarity}
Any model with additional neutrino species implies the non-unitarity
of the standard leptonic mixing matrix for the three oscillation
neutrino picture. In this scenario, the three times three mixing matrix is a block of the complete mixing matrix $U^{n\times n}$, with $n$ the total number of neutrino eigenstates. Studies on the implications of the non-unitarity can be found in the literature~\cite{Schechter:1980gr, Gronau:1984ct, Nardi:1994iv,Atre:2009rg}, as well as constraints from either neutrino experiments or those coming from charge leptons~\cite{Escrihuela:2016ube,Fernandez-Martinez:2016lgt,Blennow:2023mqx}. Recent constraints from a combined analysis of short and long-baseline experiments are reported in Ref.~\cite{Forero:2021azc}. 

For the general case of 3 active neutrinos and $n-3$ heavy neutrino states, we can define the matrix $U^{n\times n}$ as compose of four submatrices
\begin{equation}
    U^{n \times n}= \begin{pmatrix}
    N & S \\ 
    V & T
    \end{pmatrix},
\end{equation}
where $N$ is the $3 \times 3$ matrix in the light-active neutrino sector, and S describes the contribution of the extra isosinglets states to the three active neutrinos.

The neutral heavy leptons effects in the active neutrino oscillation can be factorized into the N matrix as follows \cite{Escrihuela:2015wra}: 

\begin{equation}
\label{N}
    N=N^\mathrm{NP}U=\begin{pmatrix}
\alpha_{11} & 0 & 0 \\ 
\alpha_{21} & \alpha_{22} &0 \\
\alpha_{31} &  \alpha_{32} & \alpha_{33}
\end{pmatrix} U,
\end{equation}
where U is the usual leptonic mixing matrix, and $N^\mathrm{NP}$ is the matrix characterizing the unitary violation that arises when new heavy neutrino states are introduced.

Clearly, the $3\times 3$ $N$ matrix is not unitary and, in this case,
\begin{align}
\label{unitary}
N N^\dagger = I - S S^\dagger = &N^\mathrm{NP}UU^{\dagger}N^{\mathrm{NP}^{\dagger}} \notag \\ &= \begin{pmatrix}
\alpha^2_{11} & \alpha_{11}\alpha^{*}_{21} & \alpha_{11}\alpha^{*}_{31} \\ 
\alpha_{11}\alpha_{21} & \alpha^2_{22}+|\alpha_{21}|^2 & \alpha_{22}\alpha^{*}_{32}+\alpha_{21}\alpha^{*}_{31}\\
\alpha_{11}\alpha_{31} & \alpha_{22}\alpha_{32}+\alpha_{31}\alpha^{*}_{21} & \alpha^2_{33}+|\alpha_{31}|^2+|\alpha_{32}|^2 
\end{pmatrix}.
\end{align}

The $\alpha$ parameters are related to the mixings $\cos \theta_{ij}$ and $\sin \theta_{ij}$ as~\cite{Escrihuela:2016ube}
\begin{align}
\alpha_{11} &=c_{1n}c_{1n-1}c_{1n-2}\cdots c_{14}, \notag \\
\label{diagonal terms}
\alpha_{22}&=c_{2n}c_{2n-1}c_{2n-2}\cdots c_{24}, \\
\alpha_{33}&=c_{3n}c_{3n-1}c_{3n-2}\cdots c_{34}, \notag    
\end{align}
where $\theta$ is the oscillation angle, $c_{ij}=\cos \theta_{ij}$. The non-diagonal terms are~\cite{Escrihuela:2016ube}
\begin{align}
    \alpha_{21} &= c_{2n}c_{2n-1}\cdots c_{25}\eta_{24}\bar{\eta}_{14}+c_{2n}\cdots c_{26} \eta_{25}\bar{\eta}_{15}c_{14}+\cdots + \eta_{2n}\bar{\eta}_{1n}c_{1n-1}c_{1n-2}\cdots c_{14}, \notag \\
    \label{off diagonal terms}
    \alpha_{32} &= c_{3n}c_{3n-1}\cdots c_{35}\eta_{34}\bar{\eta}_{24}+c_{3n}\cdots c_{36} \eta_{35}\bar{\eta}_{35}c_{24}+\cdots + \eta_{3n}\bar{\eta}_{2n}c_{2n-1}c_{2n-2}\cdots c_{24}, \\
    \alpha_{31} &= c_{3n}c_{3n-1}\cdots c_{35}\eta_{34}\bar{\eta}_{14}c_{24}+c_{3n}\cdots c_{36} \eta_{35}c_{25}\bar{\eta}_{15}c_{14}+\cdots\notag \\  & + \eta_{3n}c_{2n}\bar{\eta}_{1n}c_{1n-1}c_{1n-2}\cdots c_{14}, \notag
\end{align}
with $\eta_{ij}=\sin \theta_{ij} e^{-i\delta_{ij}}$, where $\delta_{ij}$ is the CP phase associated to $\theta_{ij}$. Also, the non-diagonal parameters are related with the diagonal ones through the triangle inequality \cite{Escrihuela:2016ube}:
\begin{equation}
  \label{eq:triangle}
    \alpha_{ij} \leq \sqrt{(1-\alpha_{ii}^2)(1-\alpha_{jj}^2)}.
\end{equation}
As a consequence of the non-unitarity, the oscillation probability will change.
The new oscillation probability is \cite{Escrihuela:2015wra,Miranda:2020syh}: 
\begin{align}
\label{general}
 P_{\alpha \beta}=\sum^3_{i,j}N^{*}_{\alpha i}N_{\beta i}N_{\alpha j}N^{*}_{\beta j} &-4 \sum^3_{j>i}\operatorname{Re}\left[N^{*}_{\alpha j}N_{\beta j}N_{\alpha i}N^{*}_{\beta i}\right]\sin^2 \left(\frac{\Delta m^2_{ji}L}{4E_{\nu}}\right) \\
 & +2 \sum^3_{j>i}\operatorname{Im}\left[N^{*}_{\alpha j}N_{\beta j}N_{\alpha i}N^{*}_{\beta i}\right]\sin \left(\frac{\Delta m^2_{ji}L}{2E_{\nu}}\right). \notag
\end{align}

In this work, we focus on the analysis of non-unitarity formalism in the FASER experiment. In FASER, the typical energy is in the range of $100-1000\,\mbox{GeV}$ and the distance between the source and the detector is $L=480\,m$. Therefore, the wavelength is enough to consider that our results are in the regime of a short baseline. Also, as a good approximation, we can work in the  so-called zero-distance approximation. The oscillation probability in this zero-distance case is  
\begin{equation}
\label{non-unitary}
    P_{\alpha \beta}=\sum^3_{i,j}N^{*}_{\alpha i}N_{\beta i}N_{\alpha j}N^{*}_{\beta j},
\end{equation}
where Greek letters refers to the lepton-flavor index and Latin letters denote the mass state index. In terms of the non-unitary parameters $\alpha_{ij}$ from Eq.~\eqref{N}, the oscillation probabilities in this approximation are
\begin{eqnarray}
\nonumber    P_{\mu e}&=& \alpha^2_{11}|\alpha_{21}|^2, \\ 
\label{P_NU}    P_{e \tau}&=& \alpha^2_{11}|\alpha_{31}|^2, \\
\nonumber    P_{\mu \tau} &\approx& \alpha^2_{22}|\alpha_{32}|^2, \\
\nonumber    P_{e e} &=&\alpha^4_{11}, \\ 
\nonumber    P_{\mu \mu}&=& (|\alpha_{21}|^2+\alpha^2_{22})^2, \\
\nonumber    P_{\tau \tau} &=&(|\alpha_{31}|^2+\alpha^2_{32}+\alpha^2_{33})^2.
\end{eqnarray}
\section{Experiment description and analysis procedures}
\begin{figure}[t]
	\centering \includegraphics[scale=0.6]{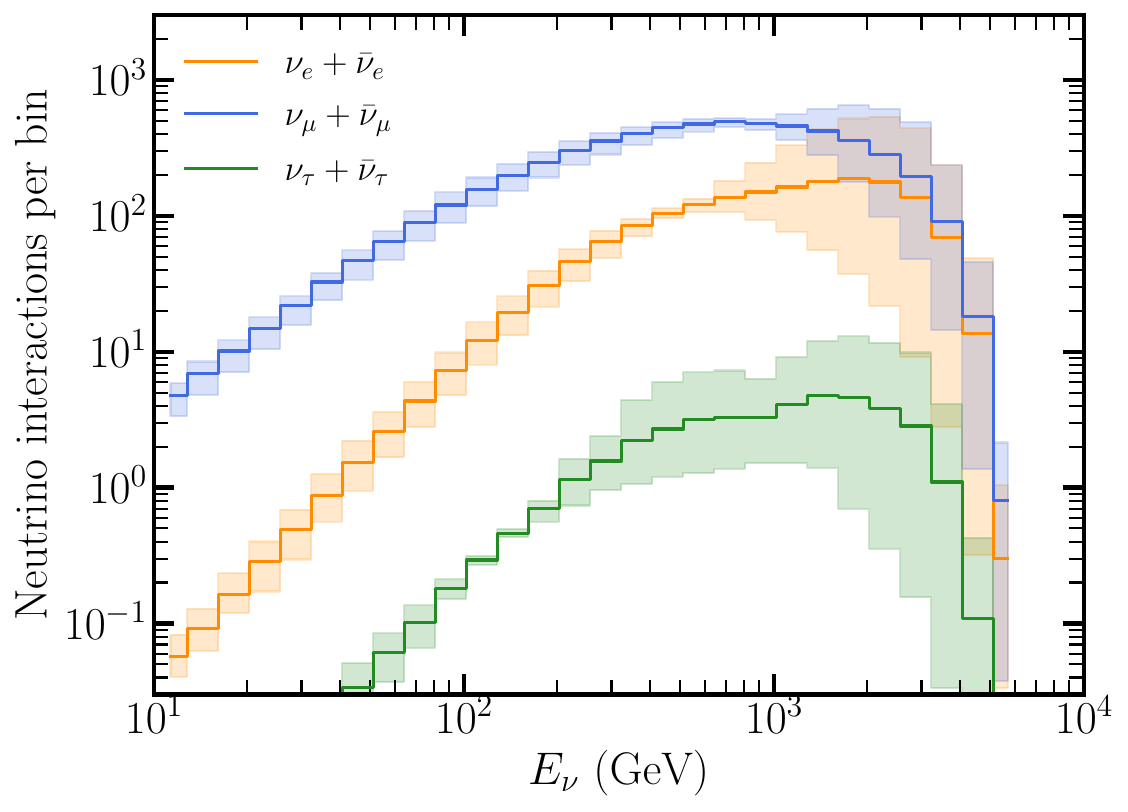}
	\caption{Expected charged-current neutrino interactions  at the FASER$\nu$ detector with 150 fb$^{-1}$ integrated luminosity, as a function of neutrino energy (recomputed from Ref.~\cite{Kling:2021gos}).}
	\label{fig:interactions}
\end{figure}

FASER$\nu$ experiment will provide an abundant neutrino flux with
thousands of expected events in the very near future. The first few
neutrino events have already been recorded by FASER$\nu$
~\cite{FASER:2023zcr}. This new experiment at the LHC opens a new
opportunity to study the non-unitarity of the neutrino oscillation
matrix in the search for heavy neutrino states due to the high
statistics for neutrino events. Being a high energy neutrino flux that
spans from 100 GeV to 1 TeV~\cite{FASER:2019dxq}, the momentum
transfer in this fix target experiment is expected to be in the range
of $Q^2\sim(10~\mbox{GeV})^2$, allowing an indirect test for new heavy
states through the non-unitary of the leptonic mixing matrix. The
FASER$\nu$ experiment has a total tungsten target mass of $1.2$~tons
and a baseline of $480$~m. The
FASER$\nu$ collaboration measures these events by measuring the
charged current (CC) with high accuracy.  The neutral-current (NC)
events are more complicated to measure due to the absence of charged
leptons in the final states, although there are some attempts to
describe the neutral-current (NC)
interactions~\cite{Ismail:2020yqc}. Although the FASER$\nu$
  collaboration has estimated the number of SM neutrino interactions
  at the detector~\cite{FASER:2019dxq}, a more recent prediction of
  these events and their uncertainties using various event generators
  has been made~\cite{Kling:2021gos}. Fig.~\ref{fig:interactions}
  shows the expected neutrino interactions at the FASER$\nu$ detector
  for each flavor. This figure was recomputed$^1$ using the information given in Ref.~\cite{Kling:2021gos} and coincides with the correponding figure of this reference. We will use
    this prediction in the following analyses on non-unitarity. 

An upgrade plan for the detection of collider neutrinos in the high luminosity era of the LHC is the FASER$\nu$2 detector~\cite{Anchordoqui:2021ghd}. With a mass of 20 tonnes and 20 times the luminosity of its predecessor, it will be able to detect two orders of magnitude more events than FASER$\nu$. Ref~\cite{Kling:2021gos} has also estimated the number of interactions at this detector.\footnote{\url{https://github.com/KlingFelix/FastNeutrinoFluxSimulation} }

In this work, we will use the zero-distance approximation to have a forecast on the sensitivity of FASER$\nu$ and FASER$\nu$2 to non-unitary $\alpha$ parameters. In this analysis, we will use 3 observables: electron, muon, and tau neutrino events. 
The SM expected events can be computed as
	\begin{equation}
		N_\alpha^\mathrm{SM} =  \varepsilon_{\alpha} N_T \int f(E_\mathrm{reco}) R(E_\mathrm{reco}, E_\nu) \, \sigma_\alpha (E_\nu) \, \phi_\alpha \, dE_\nu \, dE_\mathrm{reco}
	\end{equation}
	where $\phi_\alpha$ is the expected flux at the detector, $\sigma_\alpha$ is the neutrino-nucleus DIS cross section, $R(E_\mathrm{reco}, E_\nu)$ is a gaussian smearing function of width $0.3 E_\nu$,  $f(E_\mathrm{reco})$ it is the vertex reconstruction efficiency (taken from Fig. 9 of Ref.~\cite{FASER:2019dxq}), $\varepsilon_{\alpha}$ is the charged-lepton identification efficiency ($\varepsilon_e = 100\%$, $\varepsilon_\mu= 86\%$, $\varepsilon_\tau = 76\%$), and $N_T$ is the number of targets in the detector. 
	To estimate the number of events at both FASER$\nu$ and FASER$\nu$2 detectors, we take the estimated interactions from Ref.~\cite{Kling:2021gos} and apply smearing,  vertex reconstruction and charged-lepton identification efficiencies.
	Our estimated number of events for the complete neutrino energy range ($10^2 - 10^4$ GeV), along with their uncertainties, are shown in Table~\ref{events}.
As can be seen from Fig.~\ref{fig:interactions}, the uncertainties on the number of interactions at the detector are high, and come mostly from flux estimations. We propose a scenario where only interactions between $100-600$ GeV are taken into account in order to reduce the systematic error significantly. The expected number of events in this energy regime is also shown in Table~\ref{events}.

We compute the expected sensitivity through a $\chi^2$ analysis:
\begin{equation}
  \label{eq:chi2}
  \chi^{2}=\sum_{\alpha=e}^{\tau} \frac{(N^\mathrm{NU}_{\alpha}-N^\mathrm{exp}_{\alpha})^2}{\sigma_\alpha^{2}} + \sum_{ij}\frac{(\alpha_{ij}-\delta_{ij})^2}{\sigma^2_{ij}},
\end{equation}
where $N_\alpha^\mathrm{exp}$ is the expected measured number of events per
neutrino flavor, $N^\mathrm{NU}_{\alpha}$ is the events
number computed when non-unitarity is present, $\alpha$ refers to the
lepton flavor, and $\sigma_\alpha$ is the total
expected error (statistical and systematic). Regarding the systematic uncertainties, for FASER$\nu$, we symmetrized this
error as
an approximation, whereas for  FASER$\nu$2 we consider two scenarios, 5\% and 10\%, motivated
by the expected improvement in the flux estimation by the time the HL-LHC starts taking data.
To make a complete analysis considering the three observables that
FASER$\nu$ expects to measure, we have to consider both appearance and
disappearance channels. Therefore, the complete theoretical prediction
will depend on six different non-unitary parameters. Therefore, the
complete expressions will have more parameters than FASER$\nu$ observables,
and we need to consider priors to perform our analysis.
In Eq.~(\ref{eq:chi2}), we have included priors to the values of
$\alpha_{ij}$ that will be marginalized in our fit, using as errors,
$\sigma_{ij}$, the constraints reported in
Ref.~\cite{Forero:2021azc}. Notice that these constraints were
presented at $90$~\% C.L., then our results can be considered as
conservative.

The theoretical expected number of events will be then expressed as
\begin{align}
  N^\mathrm{NU}_\alpha &= \frac{1}{\alpha^2_{11}(\alpha^2_{22}+|\alpha_{21}|^2)}\big(N_{\alpha}^\mathrm{SM}P_{\alpha \alpha} + \sum_{\beta\neq\alpha}P_{\alpha\beta}N_{\beta}^\mathrm{SM}\big),
\end{align}
where $P_{\alpha\alpha}$ and $P_{\alpha\beta}$ are defined in Eq.~(\ref{P_NU}) and depend on the non-unitary parameters, $N_\alpha^\mathrm{SM}$ is the standard model predicted number of events for the flavor $\alpha$, and there is no sum over $\alpha$. The prefactor in the right-hand side of this equation corresponds to the correction due to the measurement of the Fermi constant, $G_F$~\cite{Langacker:1988ur,Nardi:1994iv,Atre:2009rg}, that comes from muon decay and in the case of non-unitary must be considered as $G_F=G_\mu / \sqrt{\alpha^2_{11}(\alpha^2_{22}+|\alpha_{21}|^2)}$, with $G_\mu$ as the Fermi constant measured in  muon decay. 

Since we consider several non-unitary parameters at a time,
it may happen that the disappearance events compensate for the
appearance events and no effect would be visible. Therefore, to take
into account the combined effect of the six different parameters,
considering that we have only three observables (the three neutrino
flavors), we will consider only one free parameter at a time by marginalizing over the other five parameters. 
It is important to remark that in the computation of $\chi^2$, we take into account the triangle inequality condition for all the off-diagonal $\alpha_{ij}$ parameters given in Eq.~(\ref{eq:triangle}). In other words, we have included three triangle inequality conditions.

\begin{table}[]
	\centering
	\begin{tabular}{|c|c|c|c|c|}
		\hline
		 & \multicolumn{2}{c|}{FASER$\nu$} & \multicolumn{2}{c|}{FASER$\nu$2} \\ \hline \hline
		Lepton flavor & $10^2-10^4$ GeV & $100-600$ GeV & $10^2-10^4$ GeV & $100-600$ GeV \\ \hline \hline
		$e$      & 1095$\pm$937        & 307$\pm$101    &   44230     &  20775          \\ \hline
		$\mu$   & 2807$\pm$909         &1163$\pm$190    &   193630   &  85044        \\ \hline
		$\tau$   & 19$\pm$19           & 6$\pm$4     &    767   &  314       \\ \hline
	\end{tabular}
	\caption{Expected number of events and systematic uncertainties used in the analysis, for different neutrino energy ranges. For the projections of FASER$\nu$2, we will consider two systematic uncertainty scenarios,  5\% and 10\% (see text for details).}
	\label{events}
\end{table}

\section{results}

\begin{figure}[t]
	\centering
	\includegraphics[width=0.45\textwidth]{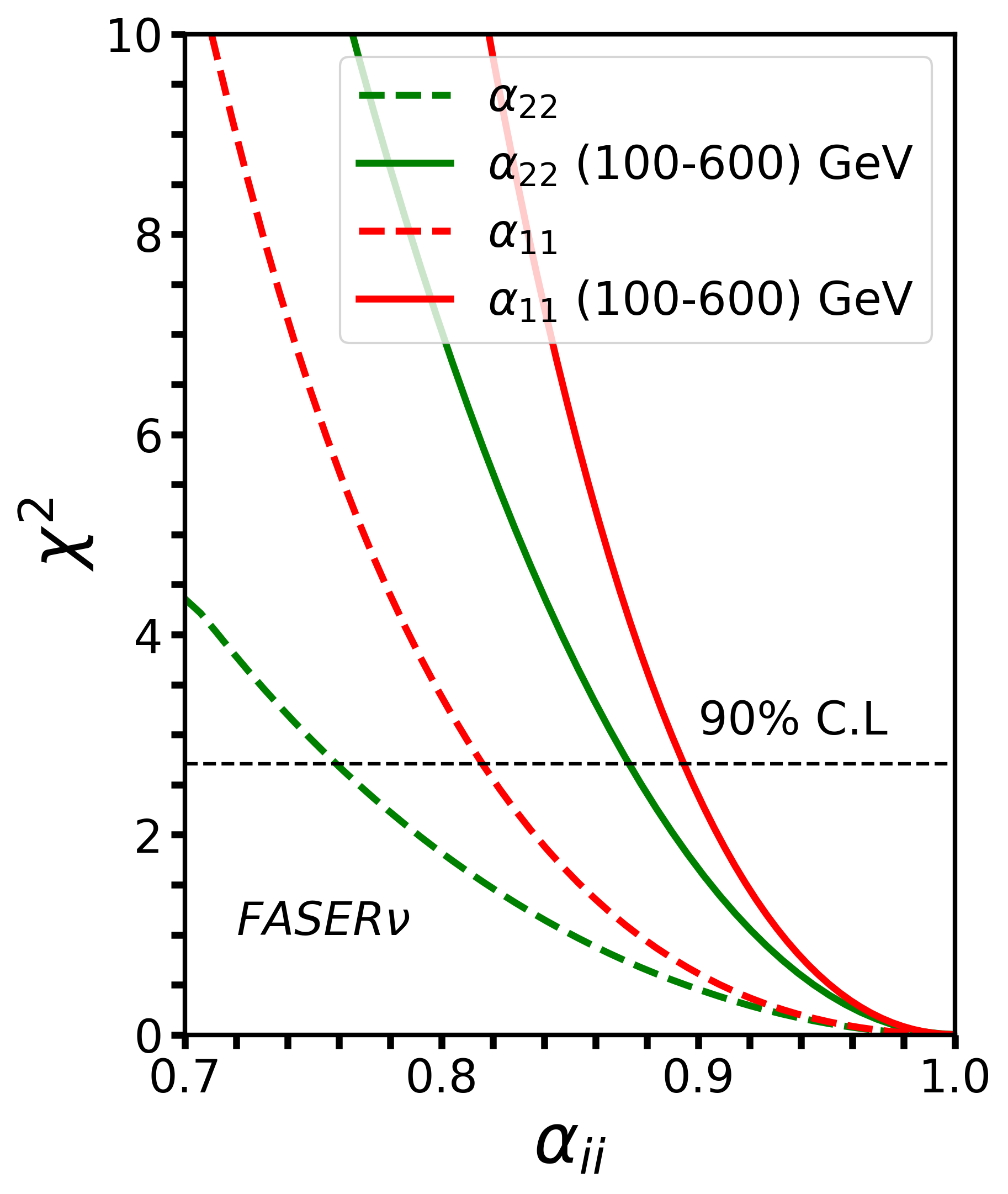}
	\includegraphics[width=0.45\textwidth]{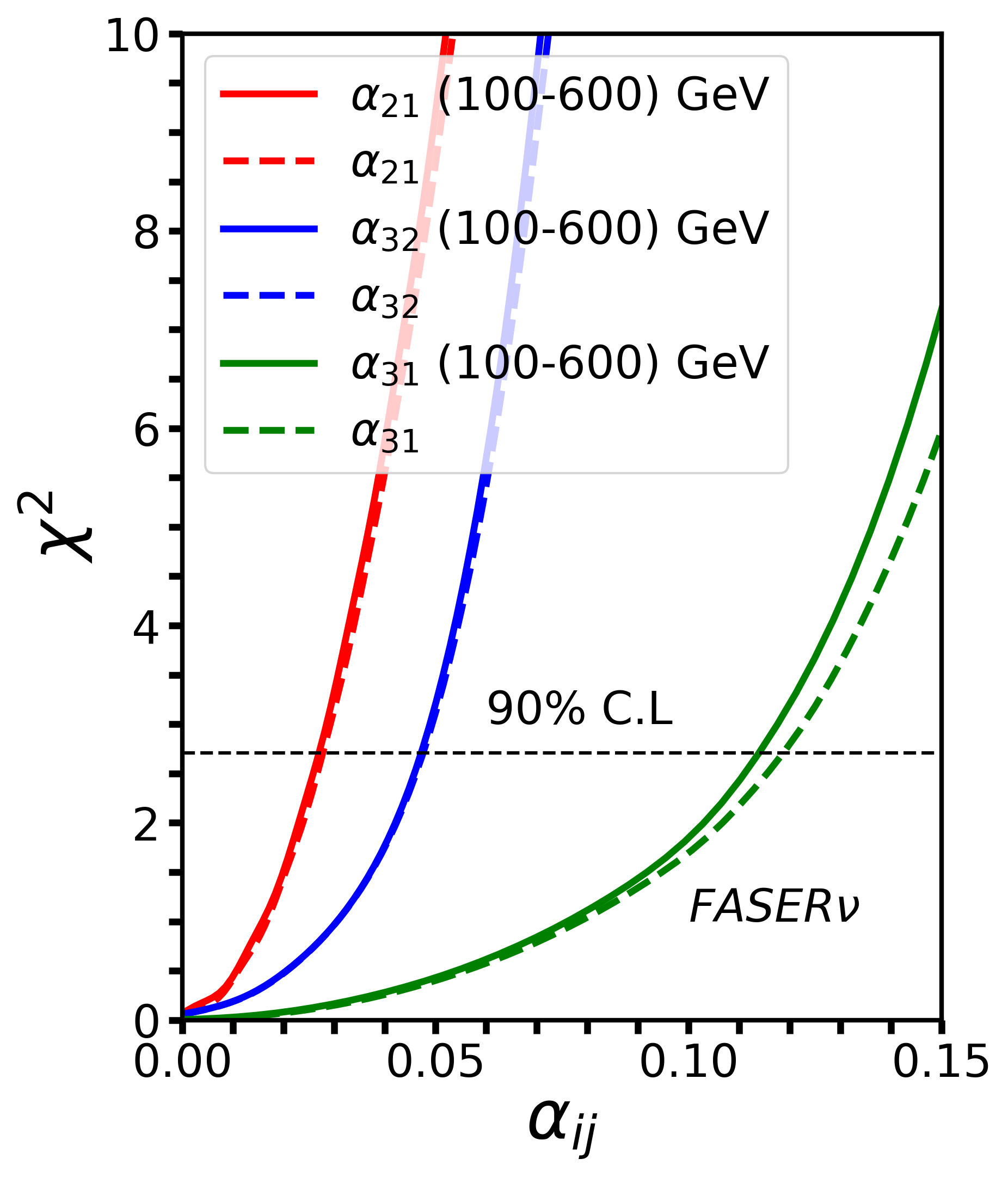}
	\caption{Sensitivity to one at a time diagonal (left panel) and non-diagonal (right panel) non-unitarity parameters for FASER$\nu$. The dashed curves represent the sensitivity for the full FASER$\nu$ energy regime, while the solid curves represent the scenario with events only between 100-600 GeV. The horizontal line shows the 90\% C.L. 
		Besides marginalization over the other non-unitary parameters, the  triangle inequality conditions have also been taken into account.}
	\label{fig:figure_1}
\end{figure}

\begin{figure}[t]
	\centering
	\includegraphics[width=0.45\textwidth]{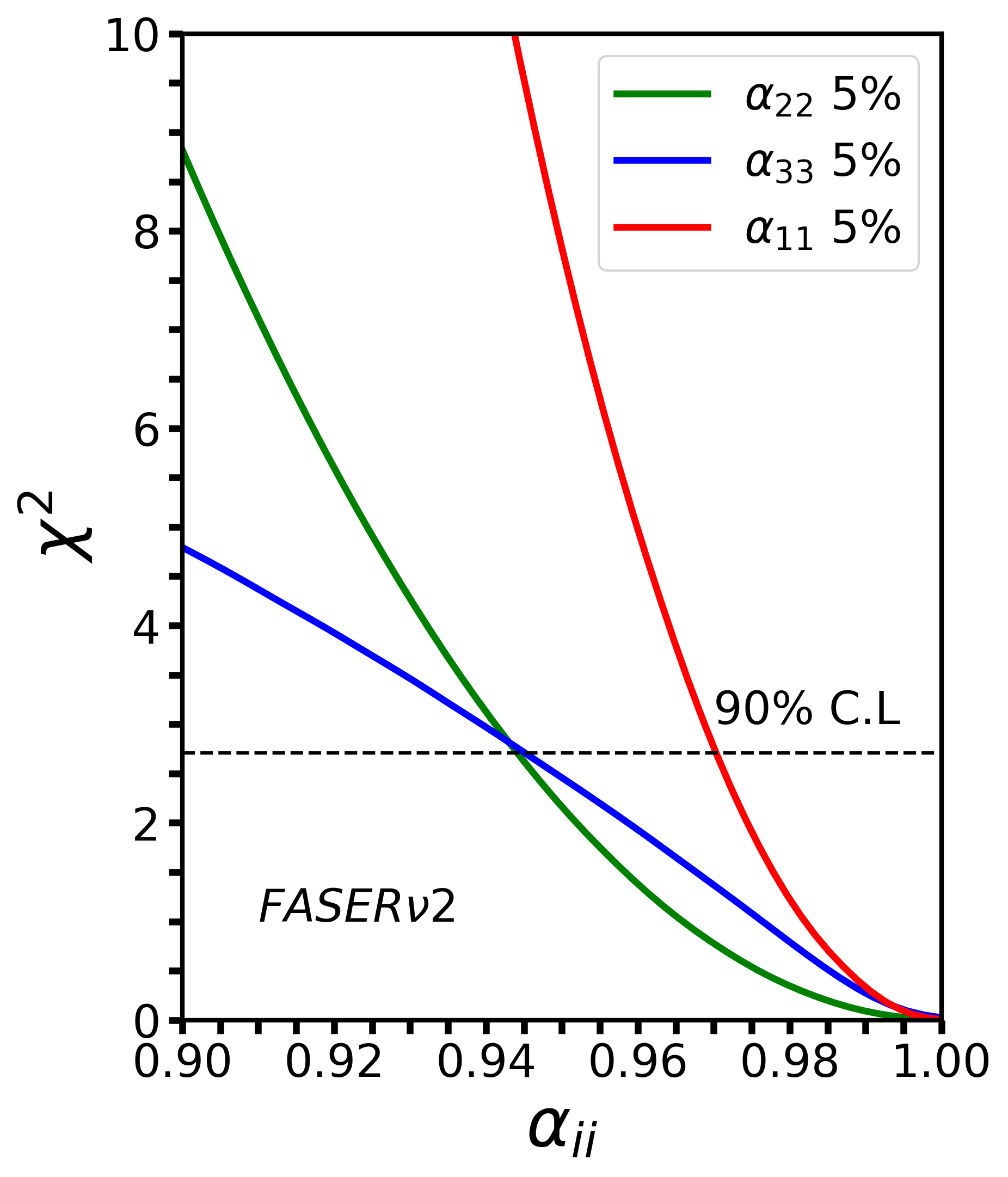}
	\includegraphics[width=0.45\textwidth]{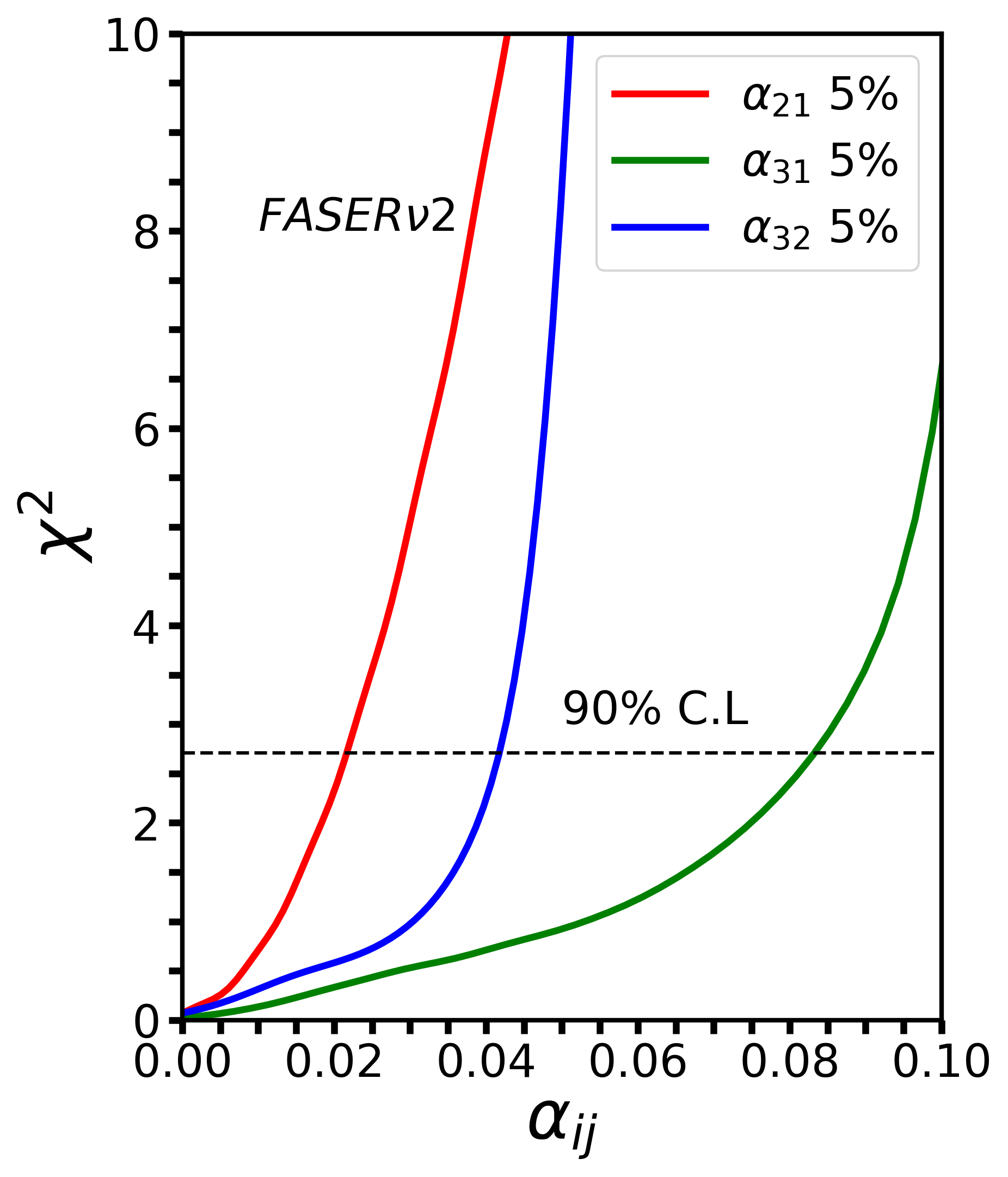}
	\includegraphics[width=0.45\textwidth]{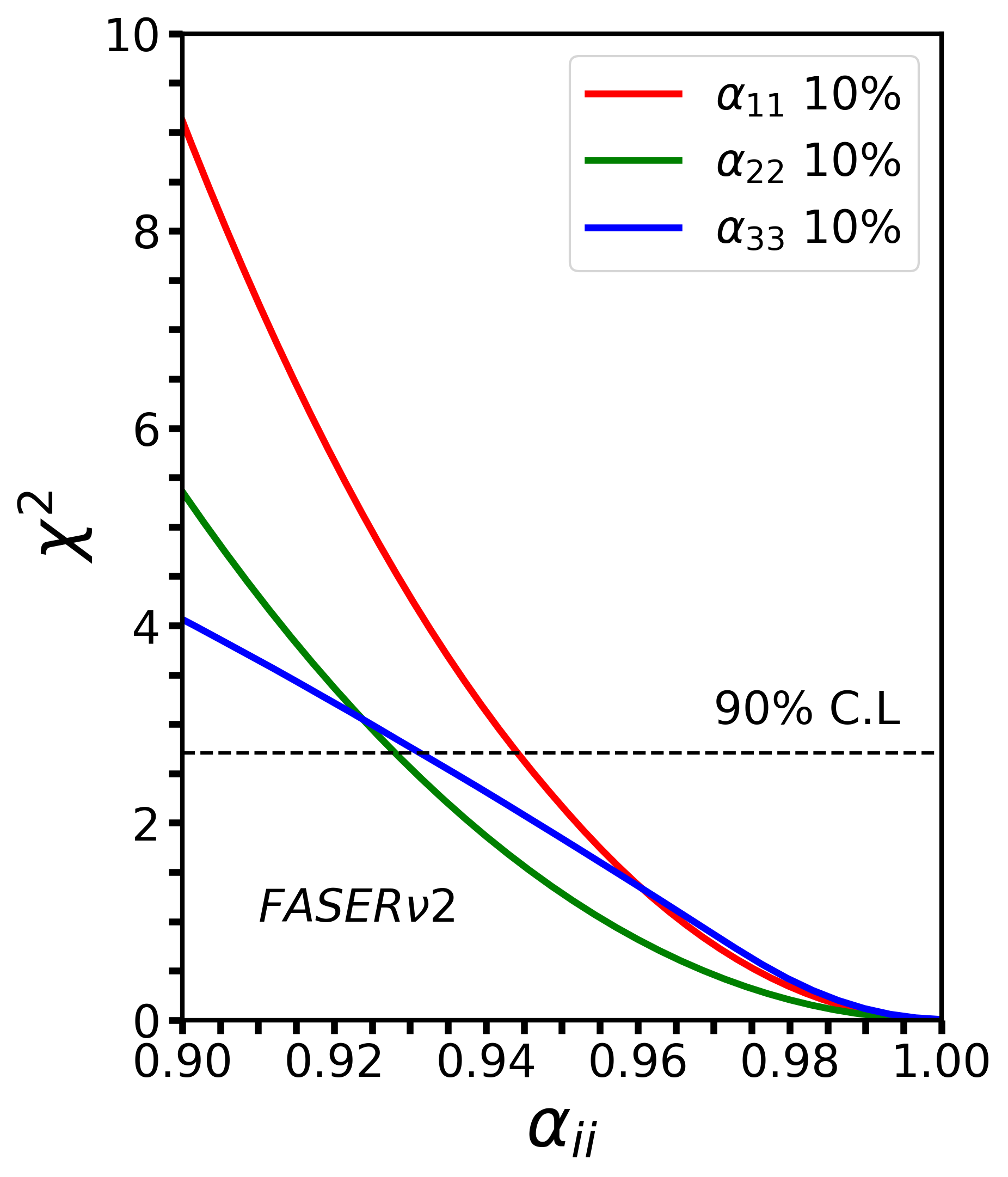}
	\includegraphics[width=0.45\textwidth]{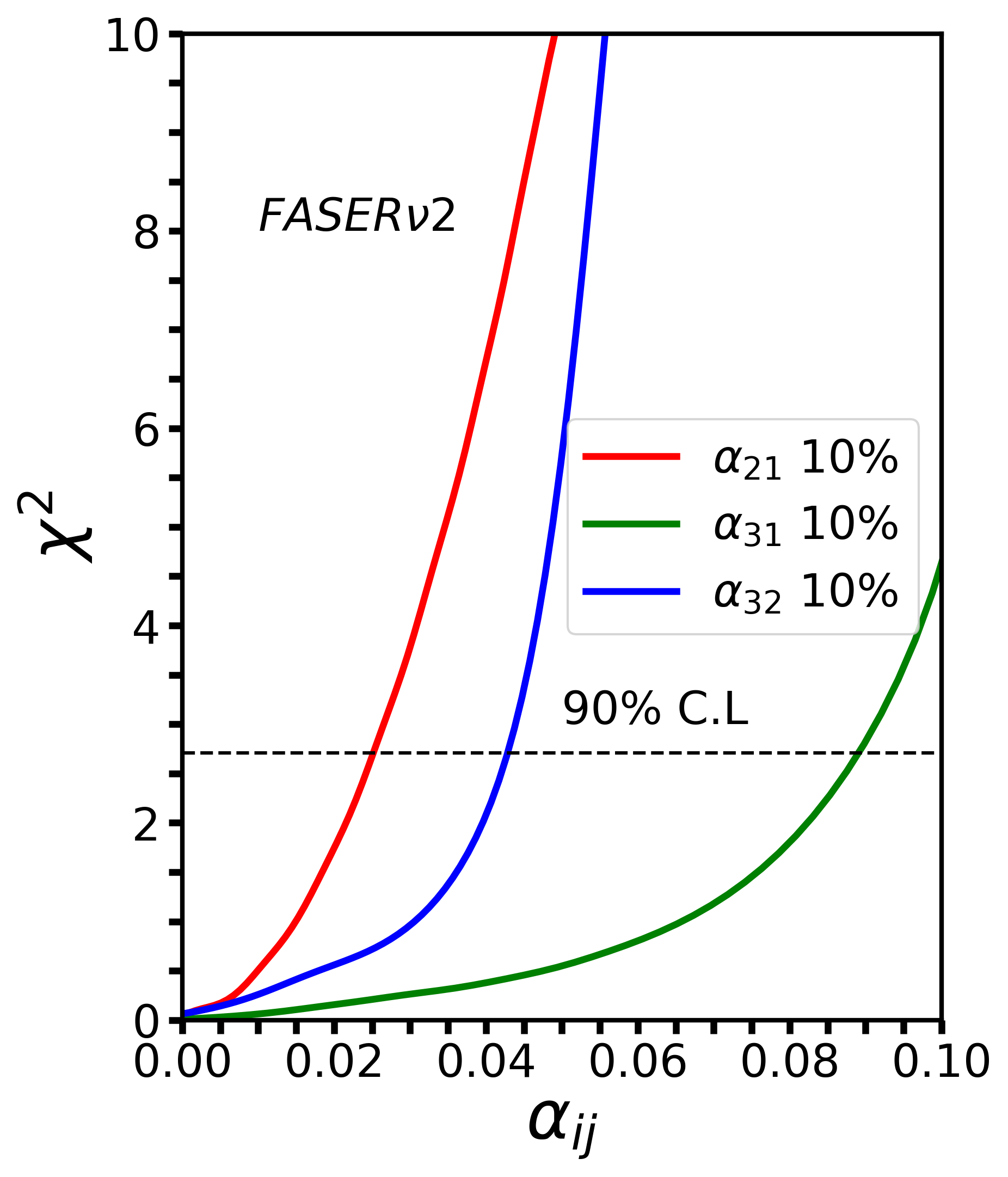}
	\caption{Sensitivity to one at a time diagonal (left panels) and non-diagonal (right panels) non-unitarity parameters for FASER$\nu$2. The upper (lower) panels correspond to the case where the systematic uncertainty is taken as 5\% (10\%). The horizontal line shows the 90\% C.L. 
		Besides marginalization over the other non-unitary parameters, the  triangle inequality conditions have also been taken into account.}
	\label{fig:figure_2}
\end{figure}

We will show in this section the results of computing the expected sensitivity for FASER$\nu$ and FASER$\nu$2 in two different energy windows. In Fig. \ref{fig:figure_1} we illustrate the expected FASER$\nu$ sensitivity for the two energy regimes already mentioned. As discussed in the previous sections, the energy range from $100-600$~GeV has smaller uncertanties. Therefore, besides the case with the full energy range, we also consider this reduced region.
However, for FASER$\nu$2, the analysis has been performed in the energy range of $100-600$ GeV because we consider systematic uncertainties very significant beyond this energy range. Results are shown in Fig. \ref{fig:figure_2}.
We must remember that this analysis considers every appearance and
disappearance channel for all neutrino flavors, as well as priors from the current limits
on $\alpha_{ij}$, hence it provides a realistic and useful projection of the FASER$\nu$ and FASER$\nu$2 capabilities to constrain the non-unitary
parameters. A summary of the expected $90$~\%  C.L. sensitivity to each
parameter is shown in Table~\ref{sensitivity}. From this table, and from Fig.~\ref{fig:figure_1}, we can
notice that it is not expected that FASER$\nu$ could improve the current
limits on non-unitary parameters. Nevertheless, for FASER$\nu$2 we found competitive results, mainly for $\alpha_{11}$ and $\alpha_{33}$. On the other hand, for the case of FASER$\nu$2, we illustrate in Fig.~\ref{fig:figure_2} how the sensitivity to non-unitarity can play a role in future global analysis, especially if we restrict ourselves to the preferable energy window that goes from $100-600$~GeV and if FASER$\nu$2 can keep under control its systematic uncertainties. Since the experiments plans to collect high statistics events (below $1$~\%) it is reasonable to expect an important campaign to reduce systematic effects. From Table~\ref{sensitivity} we can see that the most promising sensitivities are expected for the case of $\alpha_{11}$ and, especially, $\alpha_{33}$, where the tau neutrino FASER$\nu$2 events will represent a window of opportunity to shed light on this parameter.


\begin{table}[]
	\centering
	\begin{tabular}{|c | c | c | c|c|c|}
		\hline
		 & \multicolumn{2}{c|}{FASER$\nu$}  &  \multicolumn{2}{c|}{FASER$\nu$2}  &   \\ \hline \hline
	      Parameter   & $10^2-10^4$ GeV & $100-600$ GeV &  $100-600$ GeV (5\%)&  $100-600$ GeV (10\%)& Current limit   \\ \hline \hline
		$\alpha_{11} \geq$		& 0.818  	& 0.894 	& 0.970 	& 0.944& $0.969$ \\ 
		$\alpha_{22} \geq $		& 0.760 		& 0.873 	& 0.944 	& 0.928 & $0.995$ \\ 
		$\alpha_{33} \geq $		& -- 		& -- 			& 0.945   & 0.932 & $0.890$ \\ 
		$\alpha_{21} \leq$		& 0.028  & 0.027 & 0.022 	& 0.025 & $0.013$\\ 
		$\alpha_{31}\leq$		& 0.118 	& 0.114 	& 0.083 	& 0.089 & $0.033$ \\ 
		$\alpha_{32}\leq $		& 0.048 	& 0.048 & 0.042 	& 0.043 & $0.009$ \\ \hline 
	\end{tabular}
	\caption{Expected sensitivities at 90\% C.L. for all the non-unitary $\alpha_{ij}$ parameters for the FASER$\nu$ and FASER$\nu$2 experiments, for different energy ranges and systematic uncertainties. For FASER$\nu$, there is no constraint on $\alpha_{33}$.  In the last column we show the current global limits from Ref.~\cite{Forero:2021azc}. 
	}
	\label{sensitivity}
\end{table}

\section{Conclusions}
In this work, we analyzed the non-unitary effects in the context of
the FASER$\nu$ and FASER$\nu$2 
experiments. We used the approximation of zero distance
and performed an analysis of the expected sensitivity for all the
non-unitary parameters.  We find that the expected FASER$\nu$ sensitivity to non-unitarity test is in general poor, being best sensitive to the $\alpha_{21}$ parameter that might give a complementary information, useful perhaps in a global analysis. On the other hand, for the FASER$\nu$2 case, the perspectives are much better and the sensitivity to the $\alpha_{33}$ parameter could be quite competitive with current restrictions, thanks to the relatively large number of tau neutrino events that are expected in this detector. Besides, FASER$\nu$2 also has the possibility to give a competitive constraint on the $\alpha_{11}$ parameter. Since the expected statistic in FASER$\nu$2 is hugh, the main challenge rest in reducing the systematic uncertainties.
In summary, future measurements at FASER$\nu$2 may test the non-unitarity of the leptonic mixing angle in a different neutrino channel and energy region and may have competitive sensitivities for some of the non-unitary parameters.

\section{Acknowledgements}

This work has been partially supported by CONAHCyT research grant: A1-S-23238. The work
of O. G. M. and L. J. F. has also been supported by SNII (Sistema Nacional de Investigadoras e Investigadores, Mexico).

\bibliography{References}
\end{document}